# The SLICE, CHESS, and SISTINE Ultraviolet Spectrographs: Rocket-borne Instrumentation Supporting Future Astrophysics Missions

Kevin France[†,‡,¤], Keri Hoadley[†], Brian T. Fleming[†,*], Robert Kane[†], Nicholas Nell[†], Matthew Beasley[§], and James C. Green[‡]

[†] *Laboratory for Atmospheric and Space Physics, University of Colorado, Boulder CO 80309, USA, kevin.france@colorado.edu*
[‡] *Center for Astrophysics and Space Astronomy, University of Colorado, Boulder CO 80309, USA*
[§] *Planetary Resources, Inc., 93 S Jackson St #50680, Seattle, Washington 98104-2818, USA*
[*] *NASA Nancy Grace Roman Fellow*

NASA's suborbital program provides an opportunity to conduct unique science experiments above Earth's atmosphere and is a pipeline for the technology and personnel essential to future space astrophysics, heliophysics, and atmospheric science missions. In this paper, we describe three astronomy payloads developed (or in development) by the Ultraviolet Rocket Group at the University of Colorado. These far-ultraviolet (100 − 160 nm) spectrographic instruments are used to study a range of scientific topics, from gas in the interstellar medium (accessing diagnostics of material spanning five orders of magnitude in temperature in a single observation) to the energetic radiation environment of nearby exoplanetary systems. The three instruments, SLICE (Suborbital Local Interstellar Cloud Experiment), CHESS (Colorado High-resolution Echelle Stellar Spectrograph), and SISTINE (Suborbital Imaging Spectrograph for Transition region Irradiance from Nearby Exoplanet host stars) form a progression of instrument designs and component-level technology maturation. SLICE is a pathfinder instrument for the development of new data handling, storage, and telemetry techniques. CHESS and SISTINE are testbeds for technology and instrument design enabling high-resolution (R > 10^5) point source spectroscopy and high throughput imaging spectroscopy, respectively, in support of future Explorer, Probe, and Flagship-class missions. The CHESS and SISTINE payloads support the development and flight testing of large-format photon-counting detectors and advanced optical coatings: NASA's top two technology priorities for enabling a future flagship observatory (e.g., the LUVOIR Surveyor concept) that offers factors of ~50 - 100 gain in ultraviolet spectroscopy capability over the Hubble Space Telescope. We present the design, component level laboratory characterization, and flight results for these instruments.

*Keywords*: suborbital payload, spectroscopy, far-ultraviolet, interstellar medium, extrasolar planets

## 1. Introduction and Motivation

NASA's sounding rocket program has been a part of the agency's scientific portfolio since the inception of NASA in 1958. However, cosmic ray and atmospheric composition experiments (originally V-2 rockets and later transitioning to Aerobee vehicles; van Allen & Tatel 1948; van Allen & Hopfield; 1952) were being launched to study Earth's upper atmosphere several years earlier than that. The mainstay of the program has been relatively low-cost access to the space and near-space environment for direct in-situ observations (geospace and auroral physics) and access to wavelengths at which Earth's atmosphere is opaque (ultraviolet and X-ray astrophysics and heliophysics). This unique science capability is complemented by the other two pillars of the sounding rocket program: a testbed for space technology and training the next generation of scientists and engineers to lead future space endeavors.

Sounding rocket payloads enable observations at wavelengths or observing modes not accessible by orbital facilities, e.g., imaging spectroscopy at wavelengths below the $MgF_2$ reflectivity cut-off (Burgh et al. 2002; France et al. 2004; Lewis et al. 2009), wide-field near-IR imaging above the atmospheric sky-background layer (Zemcov et al. 2014), or the ability to respond to target-of-opportunity missions such as bright supernovae or comets near perihelion passage (e.g., McPhate et al. 1999). Rocket experiments also provide a platform to advance critical subsystem technology and mature their technology readiness level (TRL) through flight testing before these technologies are employed on larger, more risk averse missions. For example, spectrograph concepts, optical components, and detectors developed for suborbital payloads have matured into large scale and highly successful NASA missions, including flight verification of multi-anode microchannel arrays (MAMAs; in use on HST-STIS,

---
[¤]Corresponding author.

-ACS, and the -COS NUV mode), virtual source holographically-corrected UV gratings (Sarlin et al. 1993), and MCP delay-line detectors (Sarlin et al. 1994; Beasley et al. 2004; both enabling technology for FUSE and COS). The NUVIEWS rocket experiment (Schiminovich et al. 2001), for instance, provided a proof of concept for components of the UV imaging instrument that were later employed in the GALEX mission (Martin et al. 2005). Sounding rocket programs also can serve to expand advanced technologies developed for specific purposes or large missions into a wider astronomical application, thus capitalizing on previous technology investments (Rogers et al. 2013, Fleming et al. 2013).

The sounding rocket program is one of the few research experiences where graduate (and undergraduate) students can be involved with and gain experience in all phases of a NASA mission, from proposal through design and fabrication, to launch operations and publications. The lifecycle of a suborbital mission is three to four years, so these activities can be encompassed within the tenure of a doctoral research project. Intermediate-sized (e.g., Explorers) space missions typically have development timescales of $5 - 10$ years and the risk averse posture assumed when managing these missions prevents graduate students from getting involved with mission critical operations. As a result, students working within the suborbital program gain a more complete perspective on the requirements for completing a space mission on cost and on schedule. Students who have carried out their graduate research training in sounding rockets have a unique understanding of the relationship between science drivers and systems engineering, making them well-positioned to lead larger, orbital missions and instruments later in their careers. This is evidenced by the large number of NASA missions and instruments with PIs who trained in the sounding rocket program as graduate students, including HUT, GALEX, FUSE, HST-COS, SPEAR, CHIPS, SDO-EVE, and others.

The themes of science, technology development, and student training are echoed in all of the long-standing university suborbital programs, and in this paper, we describe the recent, current, and future ultraviolet (UV) astrophysics payloads in development at the University of Colorado. The current goal of the University of Colorado ultraviolet rocket program is to develop the technical capabilities to enable a future, highly multiplexed ultraviolet spectrograph (with both high-resolution and imaging spectroscopy modes), e.g., an analog to the successful Hubble Space Telescope-Space Telescope Imaging Spectrograph (HST-STIS) instrument, with an order-of-magnitude higher efficiency. We do this in the framework of a university program where undergraduate, graduate, and postdoctoral training is paramount, and cutting edge science investigations support our baseline technology development program.

This paper will focus on the science drivers and technological implementation of three UV spectrographs either recently launched, scheduled to be launched, or in development by the Colorado UV rocket group. The Suborbital Local Interstellar Cloud Experiment (SLICE) is a medium-resolution point source spectrograph targeting both hot ($10^5$ K) and cool ($10^2$ K) gas for absorption line studies in the $102 - 107$ nm bandpass (France et al. 2013c). SLICE was a both a science mission and an engineering development payload that provided the basis for developing and flight testing new telemetry interface and payload control computers. SLICE served as the precursor for the Colorado High-resolution Echelle Stellar Spectrograph (CHESS; France et al. 2012, Hoadley et al. 2014), an echelle spectrograph that extends the capabilities of SLICE with ~20 × the spectral resolution and ~10 × the spectral bandpass. CHESS had an initial test flight and is currently undergoing refurbishment in support of a second launch. The next rocket payload, the Suborbital Imaging Spectrograph for Transition region Irradiance from Nearby Exoplanet host stars (SISTINE) is currently in the design phase. SISTINE moves into a complementary regime of moderate resolution far-UV imaging spectroscopy: sub-arcsecond imaging combined with large bandpass and substantial field-of-view for spectrally, spatially, and temporally resolved observations of astrophysical phenomena ranging from exoplanet environment characterization to the escape of metals and molecules from low-redshift star-forming galaxies. We describe each of these payloads in greater detail below.

These payloads are also designed to respond to NASA's call for technology development and TRL maturation for future astrophysics missions. NASA's 2014 COR Technology Report lists "High-Reflectivity Optical Coatings for UV/Vis/NIR" as the highest importance goal for Cosmic Origins (Item #1 on Priority 1 list). SISTINE will provide the first test of advanced Al+LiF coatings (Quijada et al. 2014) on shaped optics and the first flight test of these coatings. Items #2 and #3 on the COR Priority 1 list both deal with large format and photon-counting UV detectors. CHESS is currently employing the new cross-strip MCPs developed as part of NASA's Strategic Astrophysics Technology (SAT) program, and we will incorporate a next generation UV-sensitive $\delta$-doped CCD detector on a future flight (also developed as part of the SAT program). The SISTINE MCP detector system will have a total active area of 220 mm × 40 mm; this large dispersion-direction dimension enables a factor of ten increase in the spectral bandpass per exposure relative to the current state-of-the-art UV imaging spectrograph (HST-STIS).

Section 2 describes a small sample of the science drivers for high-resolution point-source spectroscopy and long-slit imaging spectroscopy at UV wavelengths. In Section 3, we present the instrument-level design and component-level development work for SLICE, CHESS, and SISTINE. We describe the laboratory calibration and characterization of component level and end-to-end system performance for SLICE and CHESS in Section 4 and we conclude in Section 5 with a brief description of the flight results from SLICE and CHESS.

## 2. Scientific Drivers: From the Intergalactic to the Circumplanetary Medium

A key astrophysical theme that will drive future UV/optical space missions is the life cycle of cosmic matter, from the flow of intergalactic gas into galaxies to the formation and evolution of exoplanetary systems. A broad range of general astrophysics, tracing the lifecycle of baryons from the intergalactic medium to protoplanetary disks, requires high spectral and spatial resolution spectroscopy. A next-generation, flagship class mission will provide the aperture necessary to carry out high resolution quantitative spectroscopy of a new range of astronomical objects and provides unprecedented angular resolution for imaging spectroscopy.

### 2.1. *The Physical and Chemical Composition of the Interstellar and Intergalactic Medium: Point-source Spectroscopy with SLICE and CHESS*

The SLICE and CHESS payloads were designed to support the development of a high-resolution, low-background point source spectrograph for a future large mission (a LUVOIR Surveyor, possibly the ATLAST or HDST mission concepts; Tumlinson et al. 2015). While there are many key questions in modern astrophysics where UV spectroscopy provides unique and essential observational data, we briefly focus on a single science driver enabled by the development of SLICE and CHESS: detailed measurements of the low-redshift intergalactic medium and circumgalactic medium (IGM/CGM), focusing on the roles of feedback and accretion between star-forming galaxies and the IGM/CGM.

Primordial gas is processed in stars that explode as supernovae and inject their metals into the CGM, possibly subsequently also enriching the IGM. However, the rates and extent of this reprocessing, as well as the interaction between galaxies and the IGM are poorly understood at all redshifts. The importance of studying metal transport at low redshift cannot be overemphasized. Studies suggest that much of the IGM enrichment comes from low-mass galaxies with weak gravitational fields (Stocke et al. 2004; Keeney et al. 2006; Tumlinson et al. 2011a). Low-mass galaxies are faint ($L/L* \sim 0.1$) and even the best ground-based surveys become highly incomplete at $z > 0.1$. Key spectral diagnostics for the study of the galaxy/IGM connection reside in the rest-frame FUV (HI Ly$\alpha$, Ly$\beta$, and the Lyman series; CIV $\lambda\lambda 154.8, 155.0$; SiIV $\lambda\lambda 139.4, 140.3$; SiIII $\lambda 120.6$; and OVI $\lambda\lambda$ 103.2, 103.8 nm). COS is dramatically changing our understanding of the galaxy/IGM connection (Tumlinson et al. 2011b; Meiring et al. 2013), however, some of the most important diagnostics (Ly$\beta$, OVI, CIII) are not accessible at redshifts compatible with galaxy surveys. High ionization stages of abundant ions (particularly OVI, CIV, and NeVIII) are unique tracers of ionization mechanisms and physical conditions in accreting IGM gas and galactic outflows, and are therefore particularly important. The complex relationship between the IGM and galaxies in the present-day Universe can be addressed with a CHESS-like spectrograph onboard a future long-duration mission. Together with the Sloan Digital Sky Survey (SDSS) and optical multi-fiber spectroscopy, the CHESS technology will enable the observation of a large sample of close QSO/galaxy pairings (e.g., Tumlinson et al. 2015) in a reasonable amount of time. Figure 1 (right) shows the availability of key IGM tracers as a function of redshift.

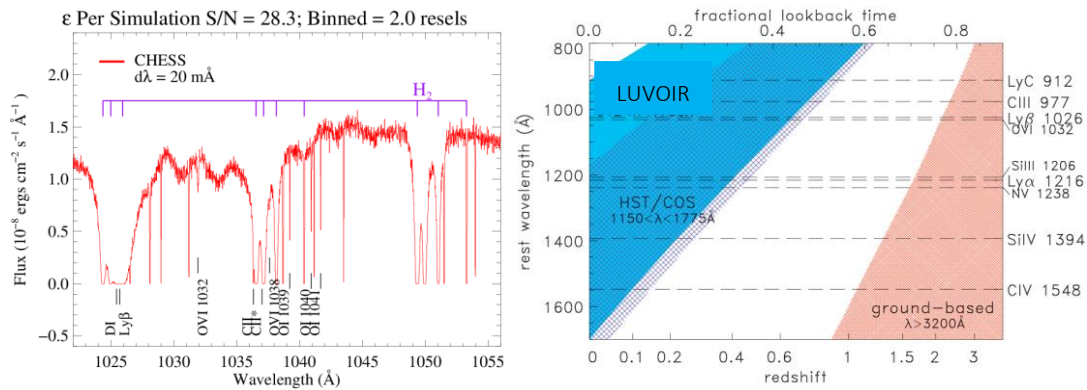

Figure 1. At left, simulated 1022-1056 Å spectrum of ε Per, the target for CHESS-2 on the 36.297 UG rocket mission, demonstrating the wealth of atomic and molecular ISM tracers observed simultaneously at high-res in the FUV. At right, accessibility to different IGM diagnostic lines as a function of lookback time. Ground-based observations (red) can only observe most IGM lines at z > 1 − 2. HST/COS (hatching) can observe Lyα at z = 0, but can't see the crucial Lyβ and O VI IGM diagnostics. To observe these, wavelength coverage at λ < 1100 Å is required (blue, figure courtesy of Charles Danforth).

The current mission of the SLICE and CHESS payloads is to connect the physical conditions of our local Galactic neighborhood to the CGM/IGM studies carried out by larger missions; the composition and physical state of the interstellar medium (ISM) is related to the cycle of massive star birth and death where by material is formed in stars, expelled into the CGM/IGM, then potentially re-accreted by the Milky Way to form future generations of stars. The local ISM (LISM) provides an opportunity to study general ISM phenomena up close and in three dimensions, including interactions of different phases of the ISM, cloud collisions, cloud evolution, ionization structure, thermal balance, turbulent motions, etc. (see review by Redfield 2006). Our immediate interstellar environment also determines the structure of the heliosphere (the momentum balance of the solar wind and the surrounding ISM). The heliosphere controls the cosmic ray flux seen in the inner solar system (Zank & Frisch 1999), which has a profound effect on the Earth, influencing cloud cover, lightning frequency, upper atmosphere chemistry, and even mutation rates of surface, deep-earth, and deep-sea organisms. The interaction between stellar winds and the ISM is a general phenomenon; all stars and planetary systems will have astrospheric interfaces. Therefore, understanding the structure of the LISM is important in evaluating the cosmic ray environment and the potential habitability of nearby exoplanets.

Because these clouds are optically thin, the distribution of ionizing sources (hot stars) determines the three-dimensional ionization structure of the LISM. Measurements of different ionization stages are required in order to probe the different ionization environments of the LISM. In addition to local ionization structure, local temperature and elemental depletion structure are also critical to understanding the three-dimensional morphology of the LISM. Determining these temperatures and abundance patterns requires high spectral resolution ($\Delta v \approx 3$ km s$^{-1}$; R ≈ 100,000; see Figure 1, left) so that contributions from thermal and turbulent motions, as well as weak lines from varying ionization states of trace species, can be distinguished. Spectroscopic measurements of the LISM at $\Delta v \leq$ 3 km s$^{-1}$ will allow us to separate the broadening mechanisms of multiple clouds on a given sightline, and it is this unique capability we will pursue with the high resolution (R ~ 120,000) CHESS payload. The paucity of strong diagnostic lines at visible wavelengths, combined with the low column densities of the nearest interstellar clouds, prevents easy observations from the ground. Because of the richness of the FUV spectral region, in terms of line densities and large oscillator strengths for the dominant species in the LISM, a space-based instrument yields the greatest science return per spectrum for a given wavelength interval. However, the HST UV spectrographs and FUSE had stringent brightness limits which prevented the observation of nearby, bright early-type stars, precisely those that would provide the most useful high-S/N data on the LISM.

## 2.2.  The Energetic Radiation Environment in the Habitable Zones Around Low-Mass Stars: Imaging Spectroscopy with SISTINE

A high-efficiency, high-angular resolution imaging spectrograph supports a wide array of astrophysical investigations (e.g., Clarke et al. 2002; Walter et al. 2003; France et al. 2010b). We focus on one particular science driver for the development of SISTINE: characterizing the critical UV irradiance spectrum of exoplanet host stars. The UV spectrum of the parent star drives and regulates the upper atmospheric heating and chemistry on rocky planets in the habitable zones (HZs; orbital radii where liquid water may be sustained for some portion of the "year") around low-mass ($M_* < M_{sol}$) stars.    At present, no stellar models can reliably predict the UV spectrum of a particular M or K dwarf without a direct observation, which prevents us from accurately modeling the spectra of Earth-like planets in these systems and evaluating their potential habitability.

Spectral observations of $O_2$, $O_3$, $CH_4$, and $CO_2$ are expected to be among the most important signatures of biological activity on planets with Earth-like atmospheres (Des Marais et al. 2002; Kaltenegger et al. 2007; Seager et al. 2009). The chemistry of these molecules in the atmosphere of an Earth-like planet depends sensitively on the strength and shape of the UV spectrum of the host star. $H_2O$, $CH_4$, and $CO_2$ are sensitive to LUV/FUV radiation ($100 - 175$ nm), in particular the bright HI Lyα line, while the atmospheric oxygen chemistry is driven by a combination of FUV and NUV ($175 - 320$ nm) radiation. The photolysis (photodissociation) of $CO_2$ and $H_2O$ by Lyα and other bright stellar chromospheric and transition region emission lines (e.g., C II λ133.5 and C IV λ155.0 nm) can produce a buildup of $O_2$ on planets illuminated by strong FUV radiation fields. Once a substantial $O_2$ atmosphere is present, $O_3$ is readily formed in multi-body reaction. $O_3$ photolysis is then driven by NUV and B-band optical photons. Therefore, on planets orbiting stars with strong FUV and weak NUV flux, a substantial $O_3$ atmosphere may arise via photochemical processes alone (Segura et al. 2005; Hu et al. 2012; Domagal-Goldman et al. 2014).

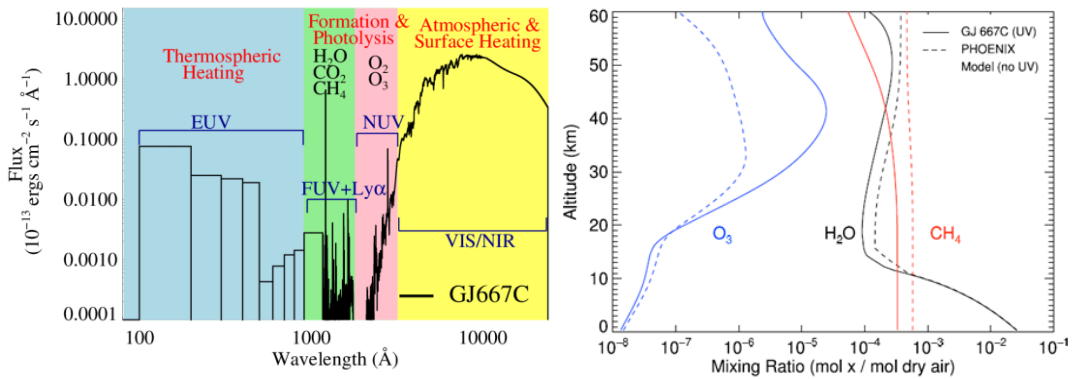

Figure 2. At left, panchromatic spectrum of the habitable zone planet-hosting M dwarf GJ 667C, illustrating the influence of each spectral bandpass on the atmosphere of an Earth-like planet orbiting this star. At right, atmospheric mixing ratios of ozone (blue), methane (red), and water (black) for an Earth-like planet orbiting at the 1 AU equivalent distance around GJ 667C (solid) and photosphere only model (dashed; figure courtesy of Sarah Rugheimer). Realistic UV irradiances predict more than an order of magnitude increase in the $O_3$ abundance.

It has been recently shown that M and K dwarf exoplanet host stars display FUV/NUV flux ratios in the range ~ $0.01 - 1$, $> 10 - 1000$ times the solar ratio (France et al. 2013a). This ratio is driven by the combination of strong Lyα emission in the FUV and relatively little NUV emission owing to the lower effective temperature of the stellar photosphere (Figure 2, left). Recent models predict that when realistic inputs for the UV radiation field are included (Figure 2, right), there will likely be detectable $O_3$ abundances on terrestrial planets in the HZ of M dwarfs without the presence of biological life (Tian et al. 2014; Gao et al. 2015). Therefore, a detailed knowledge of the stellar spectrum will be critical for interpreting observations of potential biomarkers on these worlds when they are detected in the coming decades.

No previous or current observatory offers simultaneous spectral coverage of these important spectral tracers discussed above, O VI (103.5 nm) through Lyα (121.6 nm) to C IV (155 nm). Owing to the frequent, yet aperiodic and abrupt, time variability of these sources, simultaneous observation of spectral tracers probing a range of stellar atmospheric layers is essential. Flares can alter the UV luminosity by factors of 10 on timescales of seconds, and even non-flaring states are characterized by stochastic fluctuations of ~30% minute timescales (Loyd & France 2014). Furthermore, *HST* observations typically require two instrument configurations to acquire both Lyα and C IV observations – imaging spectroscopy with STIS is required to subtract the geocoronal Lyα signal while COS observations are often required to provide sufficient signal-to-noise (S/N) of the fainter high-ion lines.

SISTINE was designed specifically to address these technical hurdles and to provide the most robust stellar irradiances for state-of the–art terrestrial atmosphere models. SISTINE provides a broad bandpass (100 − 160 nm) to simultaneously cover O VI thru C IV, employs advanced optical coatings to provide sufficient sensitivity to measure a suite of lines in a single observation, and utilizes a new imaging spectrograph design which provides the spatial resolution necessary to robustly subtract geocoronal Lyα and to resolve nearby binary exoplanet host star systems.

## 3.  Instrument Description and Component-level Development

In this section, we describe the design of each of the three instruments, and present a detailed discussion of the technical advancements pursued by each, including lessons learned during fabrication, calibration, and field-operations of the payloads.

### 3.1.  *Suborbital Local Interstellar Cloud Experiment (SLICE)*

The SLICE payload consists of a classical Cassegrain telescope followed by a modified Rowland circle spectrograph and a microchannel plate (MCP) intensified ranicon detector (Schindhelm et al. 2010; Kane et al. 2013; France et al. 2013c). A ray-trace of the instrument superimposed on a mechanical drawing is presented in Figure 3. The telescope focuses starlight onto a 50 μm × 3 mm entrance slit, the size of which was chosen to minimize the contributions from wide-field airglow emission that would contaminate the science spectrum. In a traditional Rowland spectrograph, a spherical grating disperses light from the entrance aperture into a spectrum, focused along a circle with a diameter equal to the radius of curvature of the grating. Our system uses a Horiba Jobin-Yvon fabricated aberration-correcting holography to flatten the spectral field to flatten the focal plane and improve the spectral resolution. The spectrograph is theoretically capable of a resolving power of 8,000 over the bandpass 102.0 − 107.0 nm; however, the spatial resolution of the ranicon detector ($\Delta \approx 100$ μm) limited the instrumental resolving power to approximately 5,000 (described below).

The instrumental bandpass was selected to simultaneously observe H I Lyβ, O VI (λλ 103.2, 103.8 nm), neutral metals (O I and Ar I), and multiple Lyman bands of $H_2$ and deuterated molecular hydrogen (HD). The $H_2$ and HD lines contaminate the 103.2 nm O VI feature, and therefore it is necessary to measure them separately to remove confusion from the O VI. Similarly, continuum placement is often the largest systematic source of uncertainty in deriving quantitative results from absorption line spectra; a broad enough bandpass for robust continuum determination is essential.

| Telescope | | Spectrograph | |
| --- | --- | --- | --- |
| F#: *f/7*<br>Slit Dimensions: 50µm x 3mm (7.25" x 5.13') | | Bandpass (Å): 1020 to 1070<br>Resolving Power: R = 5300<br>F#: *f/7* | |
| Primary Mirror | Secondary Mirror | Diffraction Grating | Detector |
| Eccentricity: 1<br>Radius (mm): 1730<br>OD (mm): 203.2<br>Coating:<br>Aluminum/Lithium Fluoride<br>(Al/LiF) | Eccentricity: 4.1<br>Radius (mm): 1862.5<br>OD (mm): 86.4<br>Coating: Aluminum/Lithium<br>Fluoride (Al/LiF) | Eccentricity: 0<br>Radius (mm): 998.8<br>OD (mm): 158.8<br>Coating: Aluminum/Lithium<br>Fluoride (Al/LiF)<br>Grating Ruling: Holographic<br>Groove Density: 4650<br>grooves/mm | Type: 25mm diameter Open<br>Faced MCP<br>Photocathode: Rubidium<br>Bromide (RbBr)<br>Anode: Resistive |

Table 1. Instrument specifications for the SLICE sounding rocket payload.

### 3.1.1. *SLICE Optomechanical Structure*

It is traditional in sounding rocket payloads to mount optical components and detectors to bulkheads that are spaced throughout the experiment section. This is particularly true of evacuated payloads, such as those described here, where the bulkhead can serve as both an optical bench and a vacuum barrier. In SLICE, a bulkhead between the telescope and spectrograph provides a mounting platform for interior structures (Figure 3). The telescope components are mounted to the aft side (the aft side being towards the rocket motors in launch configuration, and to the right in Fig. 3) of the bulkhead while the spectrograph components are mounted to the forward side.

The primary mirror assembly is on a three point spring mount, 2-3/4 inches from the center mechanical axis of the payload. The mirror is made out of an aluminum substrate (see below) which allows us to directly bolt to an optical bench and alignment fixtures with identical CTE. The optics are required to maintain arcsecond-level alignment during launch loads (~12 g), therefore a highly constrained all-thread, spherical washer, and nut mounting design was developed (Figure 4a). The mirror is bolted to a mounting plate at 3 points (equally spaced on a bolt circle concentric with the optical axis; Figure 4a) with spherical washers between the optic and the mounting plate. The spherical washers reduce mounting stress that might be induced by differences in angular alignment between the optic and mounting plate. A set of three spring-loaded compression bolts are included to keep the mounting plate pressed against the screw heads during adjustment and to provide return force that counteracts launch loads. The three springs are rotated 30 degrees relative to the optical mounting screws such that the load path from adjustment has the longest route to travel to cause any warping of the optic.

The secondary mirror assembly and the ST-5000 star tracker are secured at the end of two cantilevered tubes fastened to the aft side of the primary bulkhead directly over the primary mirror assembly and the detector respectively. The diffraction grating is mounted to an internal metering structure that is then mounted to the forward side of the spectrograph/telescope bulkhead. This metering structure is an aluminum tube that also acts as a heat shield to thermally separate the spectrograph from flash-heating of the rocket skins during powered ascent and acts as a light baffle for any stray light (geocoronal or astrophysical) entering through the open shutter door. The focus and alignment of the grating was controlled by adjustment nuts on a 3-point mount. In practice, the grating is adjusted during vacuum alignments; calibration data is acquired during end-to-end testing, focus is assessed by fitting the width of spectral features at each focus step (see Section 4.1). Once best focus has been realized (for SLICE, this is when we reach the detector resolution), the grating is held by locking nuts and epoxy. To give the reader a feel for the hardware, we include a photo of the SLICE grating in Figure 4b. The detector is also mounted on the aft side of the bulkhead and images the dispersed light off of the grating approximately 4.3 inches from the central mechanical axis (Figure 3).

The telescope and spectrograph are matched systems at *f/7*, with the grating slightly oversized to ensure capture of the full light beam from the telescope. The primary and secondary are separated by 500 mm, and the back focal distance from the primary vertex to the slit is 100 mm. The telescope optics were diamond-turned out of aluminum by Nu-Tek Precision Optical Corporation. After diamond-turning, the optical components were plated with nickel, diamond-turned again, and then polished. An Al+LiF overcoating was applied to the optics by the GSFC Thin Films Laboratory to enhance FUV reflectivity in the SLICE bandpass. We elected to employ a holographically-ruled diffraction grating because traditional mechanically ruled gratings are not capable of the required level of aberration control without highly aspheric substrates. The telescope and grating specifications are described in Table 1.

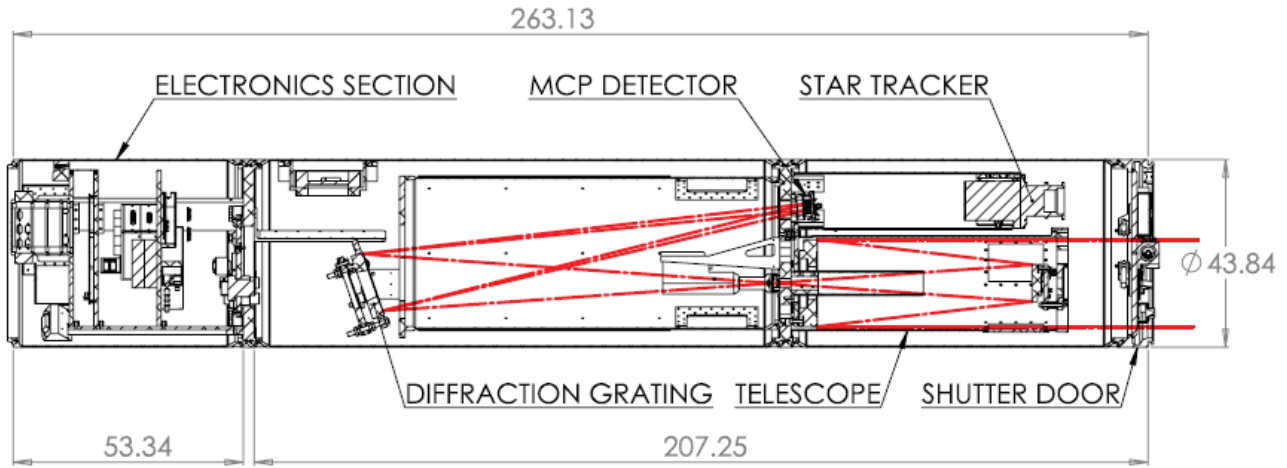

Figure 3. Schematic view of the Sub-orbital Local Interstellar Cloud Experiment (SLICE). Dimensions are in centimeters (figure adapted from France et al. 2013b). The optical elements (light enters from the right in this view) are: 1) primary mirror, 2) secondary mirror, 3) telescope focus where a 50μm × 3mm slit is cut into a mirrored slit-jaw, 4) diffraction grating, and 5) photon-counting far-UV detector. Each section of the payload is separated by bulkheads. From left to right: a vacuum bulkhead separating the electronics and the spectrograph section, the spectrograph/telescope bulkhead that serves as the instrument optical bench, and the vacuum shutter bulkhead that separates the experiment from the motor sections.

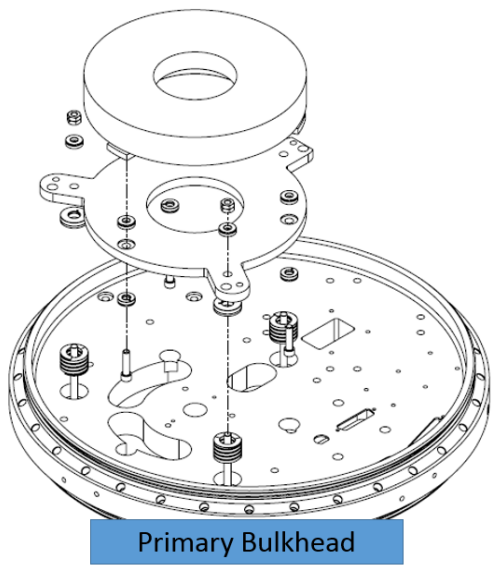
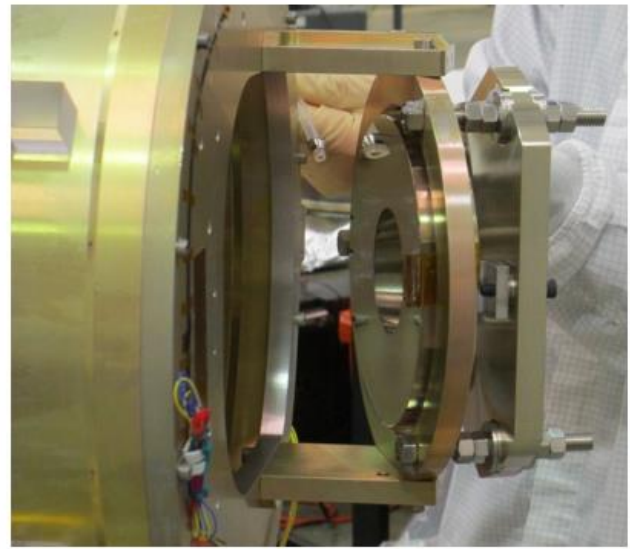

Figure 4. Optical mounts must be designed to maintain arcsecond-level pointing after experiencing launch loads in the range 10 – 20g. At left, we show a technical drawing of the SLICE primary mirror assembly. In this exploded view, the 8-inch diameter primary mirror is the top component in the drawing. The spring-loaded all-thread, spherical washer, and nut mounting design are shown along the dash-dotted vertical lines. The diffraction grating employs a three-point mounting scheme, and the grating is epoxied into place following laboratory alignment. We show a picture of the grating as installed on the end of the optics bench at right (Kane et al. 2013).

### 3.1.2. *SLICE Detector*

SLICE employs a resistive anode, 25mm diameter open-faced MCP lab detector purchased from Quantar Technologies that was ruggedized by the Colorado UV rocket group. The body of the detector head was designed to hold the MCP stack, as well as a quantum efficiency (QE) grid and an ion-repelling (IR) grid. The top plate is coated with an opaque rubidium bromide (RbBr) photocathode that produces electrons following the absorption of a photon. The electrons are accelerated down the five microchannel plates' holes through an applied high voltage. The five MCPs are arranged in a "chevron – Z" configuration. The top two MCPs are arranged so the pores of the MCP's form a chevron pattern, while the bottom three MCP's are stacked in a Z pattern, to allow for higher gain when the electron shower collides with the anode. The anode's shape allows for the charge to flow to the four corners of the anode. These signals are sent to a pre-amplifier that amplifies the pulse and combines the four signals. These are then sent to a position computer, which uses a simple algorithm to calculate the position of the electron cloud based on the charge distribution to the four corners of the anode. The detector achieves a somewhat suboptimal 100 μm spatial resolution (relative to the nominal specification of 62.5 μm) due to space constraints forcing the use of 2m preamplifier cables that add capacitive noise to the sensitive input signal. The quantum-efficiency (QE) enhancement grid is a metal mesh (70 lines/inch with 95% transmission) placed just above the top-most MCP and set at an even higher negative voltage to extend the electric field and help force the electrons back down into the pores rather than escaping out the "top". The high voltage (HV) was supplied by an Ultravolt HV power supply with integrated voltage and current monitors. The IR grid is a second a metal mesh (20 lines/inch, 95% transmission) placed in front of the QE grid and electrified to positive bus voltage in an attempt to drive away ions and keep the global background count rate low. Interestingly, global background rates were of order 150 counts/s/cm$^2$ during flight, with no detectable variation (measured during slews between different targets, see Section 5.1 below). This level is about 50 times the background rate measured during laboratory testing, and additional studies are ongoing to assess the ability of un-rejected, upper atmospheric ions to induce this elevated background.

A ruggedized PC/104 Intelligent Data Acquisition Node (IDAN®) computer from RTD Embedded Technologies, Inc. served as the SLICE flight computer. The primary function of this system is to handle the interaction between the detector electronics and NASA-supplied telemetry systems (referred to as the detector/telemetry interface or TMIF). The new PC/104 computer was part of the technology developed for the SLICE mission as a demonstration for data and event handling on future suborbital missions. These modules are all 'commercial-off-the-shelf' (COTS), allowing for a quick turnaround if one is damaged. The IDAN also provides the ability to save data on-board to a solid state hard drive. This is an important onboard computing development for future payloads as data rates increase with new detector technology to the point where current NASA telemetry systems cannot downlink all the data during flight. This feature was successfully tested on SLICE by saving a copy of the flight data on board while also telemetering it down to a ground station at White Sands Missile Range.

### 3.2. *Colorado High-resolution Echelle Stellar Spectrograph (CHESS)*

CHESS is an objective echelle spectrograph operating at $f$/12.4 and resolving power of R ≈ 120,000 over a bandpass of 100 − 160 nm. CHESS is comprised of a mechanical collimator, e-beam etched echelle grating, holographically ruled cross-disperser, and an FUV detector system. The spectrograph, detector, and supporting electronics are packaged into a standard 17.26" Wallops Flight Facility (WFF) provided rocket skin (Figure 5). The CHESS instrument has been described in previous technical conference proceedings (France et al. 2012b; Hoadley et al. 2014); we present a brief description here and a calculation of the instrumental efficiency below. The CHESS optical path can be described as follows (and see Figure 5):

- A mechanical collimator, consisting of an array of 10.74mm$^2$ × 1000mm anodized aluminum square tubes with a total collecting area of 40cm$^2$ and 0.9° field of view, prevents off-axis light from entering the spectrograph.
- An echelle grating (100 mm × 100 mm × 0.7 mm; Al+LiF-coated), with a groove density of 69 grooves mm$^{-1}$ and angle of incidence (AOI) of 67°, intercepts and disperses the stellar light. CHESS operates in orders m = 266 − 166. However, the 69 grooves mm$^{-1}$ ruling density has proven impossible to fabricate at high efficiency and we have modified the design to accommodate a 100 grooves mm$^{-1}$ ruling to increase the efficiency of the instrument. A new, 100 grooves mm$^{-1}$ e-beam etched echelle developed by JPL's MicroDevices Lab is being tested at Colorado at the time of writing.

- A holographically ruled grating (Al+LiF coated) is the cross dispersing optic. This grating is ion-etched into a laminar profile, using a deformable mirror in the recording solution, on a toroidal substrate for maximum first-order efficiency.
- A circular, 40 mm diameter cross-strip anode MCP detector (Vallerga et al. 2010) records the echellogram. This photon-counting device is capable of total global count rates around $10^6$ Hz.

| Mechanical Collimator | | Spectrograph | |
|---|---|---|---|
| FOV: 18.5' x 18.5' | | Bandpass (Å): 1000 - 1650 | |
| Collimator Dimensions (mm): 10.74 x 10.74 x 1000 | | Resolving Power: R ~ 120,000 | |
| Collecting Area (cm²): 40 | | F#: $f$/12.4 | |
| Echelle (LightSmyth) | Cross Disperser (Horiba Jobin-Yvon ) | | Detector (Sensor Sciences) |
| Shape: Flat | Shape: Toroidal | | Type: Open Faced MCP |
| Dimensions (mm): 100 x 100 x 0.7 | Radius (m): 2500.25/2467.96 | | OD (mm): 40 |
| Material: Silicon | Dimensions (mm): 100 x 100 x 30 | | Anode: Cross strip |
| Coating: Aluminum/Lithium Fluoride (Al+LiF) | Material: Fused silica | | Photocathode: Cesium Iodide (CsI) |
| Grating Ruling: Mechanical | Coating: Aluminum/Lithium Fluoride (Al+LiF) | | Global count rate (Hz): $10^6$ |
| Groove Density (gr/mm): 69 | Grating Ruling: Holographic | | Pixel format: 8k x 8k |
| Blaze angle (⁰): 67 | Groove Density (gr/mm): 351 | | Spatial resolution: 25 µm |

Table 2. Instrument specifications for the CHESS sounding rocket payload.

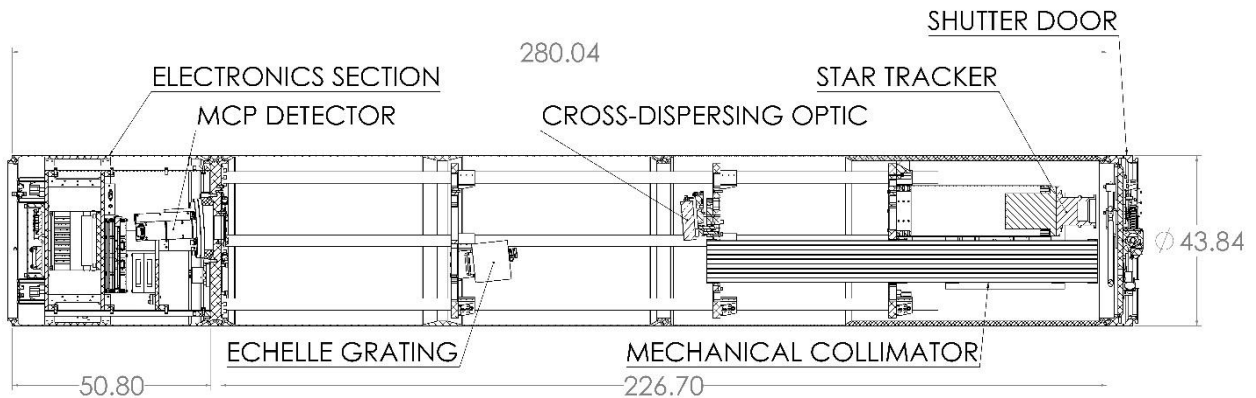

Figure 5. Schematic of the spectrograph section of CHESS. Labeled are relevant disk structures and optical components.

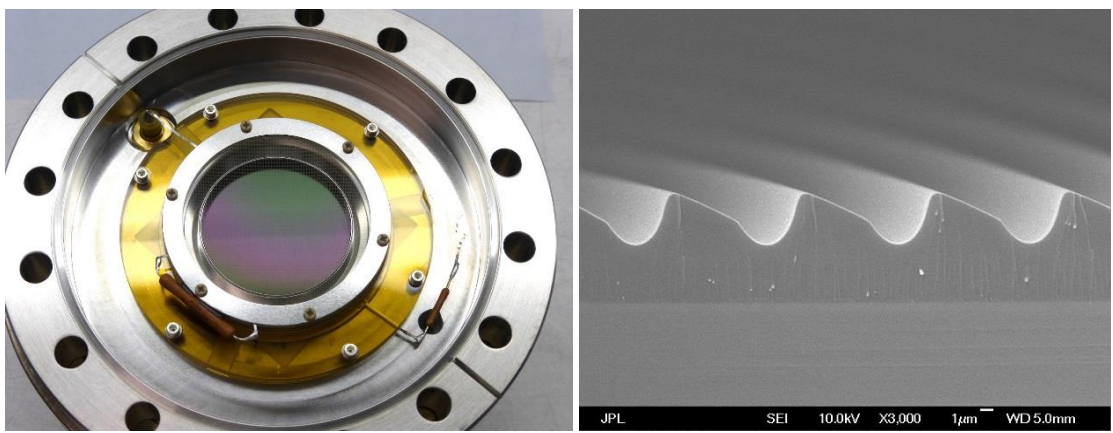

Figure 6. New hardware for the CHESS-2 payload: *(left)* the 40mm cross-strip anode MCP detector following the deposition of a new photocathode at Sensor Sciences. *(right)* Scanning Electron Microscope cross-sectional image of the electron-beam etched echelle grating produced at JPL. The 100 groove mm⁻¹ design enables higher groove efficiency, but at the cost of continuous spectral coverage at λ > 1300 Å.

### 3.2.1.  *CHESS Optomechanical Structure*

CHESS is an aft-looking payload that uses 17.26" rocket skins, and is split into two sections separated by a hermetically-sealed bulkhead. The overall length of the payload, excluding the shutter door, is 110.25" from mating surface to mating surface and the weight of the payload, including the shutter door, was measured to be 359.6 lbs. The shutter door (provided by NASA's Wallops Flight Facility) is the only moving component in the experiment section during flight. The electronics skin is 20.00" long, has three panels and one umbilical pocket for an RJ45 Ethernet port. The detector is mounted with a hermetic seal to the forward side of the 1.00" thick spectrograph/electronics section bulkhead and faces aft, into the vacuum section.

The vacuum section uses two, 44.63" long rocket skins with hermetic joints. The only mechanism is a manual, locking butterfly vacuum valve attached on the vehicle 180º line, near the aft end of the vacuum section. This allows us to evacuate the experiment throughout build-up, integration and preflight activities, safeguarding the sensitive coatings on the optics and detector. A carbon-fiber space frame is attached to the aft side of the electronics/experiment bulkhead, and suspends the star tracker, square tube collimator, echelle and cross-disperser optic in place. The space frame is comprised of three aluminum disks attached to five, 1.00" diameter x 72.00" long carbon fiber tubes.

### 3.2.2.  *CHESS Optics*

CHESS only has two primary optical components, the echelle and cross-dispersing gratings. The echelle grating is fed by a mechanical collimator, a set of 10.74mm x 10.74mm (OD) x 1m long black-anodized aluminum square tubes bonded together. These structures serve mainly to restrict the acceptance angle into the spectrograph (the effective field-of-view) and to limit the total geocoronal airglow signal (particularly the bright Lyα line) in the data.

The CHESS flight echelle for the initial flight (see Section 5.2 below) was a 100 mm x 100 mm x 0.7 mm lithographically-etched silicon wafer designed for a groove density of 69 grooves mm$^{-1}$ and AOI = 67°. Lightsmyth Inc. was a project partner for the development of a dry-etched lithography echelle grating design utilizing corner-cube groove profiles to minimize the scattered light issues that have historically plagued UV echelle gratings. Lightsmyth's experimental ruling technique was not able to cut square grooves and as a result the echelle did not meet the specified performance (peak order groove efficiency, as delivered, at 1216 Å = 1.5%; attributed to problems with achieving the desired groove density and AOI). Part of the Colorado UV rocket program's charge is research and development of multiple new technologies for UV gratings; we are currently working with JPL's Microdevices Laboratory on an electron-beam etched echelle. The JPL electron-beam results have not been appreciably better than the Lightsmyth gratings at 69 grooves mm$^{-1}$ and we are currently working with a slightly modified design that can accommodate a higher efficiency 100 grooves mm$^{-1}$ electron-beam etched grating. Test gratings have been delivered and tested (groove efficiencies > 5 × the Lightsmyth echelle flown on CHESS-1); the flight grating (to be flown on CHESS-2 in February 2016) is in production (see Figure 6, right).

The CHESS cross disperser grating is a 100mm × 100mm × 30mm fused silica optic with a toroidal surface profile. The toroidal surface shape separates the foci of the tangential and sagittal axes of the dispersed light. The optic works at $f$/12.4, focusing light spatially and sagittally on the detector. The Horiba Jobin-Yvon fabricated cross dispersing optic is a new type of imaging grating that represents a new family of holographic solutions. The line densities are low (351 lines mm$^{-1}$, difficult to achieve with the ion etching process), and the holographic solution allows for more degrees of freedom than was previously available with off-axis parabolic cross dispersing optics. The holographic ruling pattern corrects for aberrations that otherwise could not be corrected via a parallel ruling process. The grating is developed under the formalism of toroidal variable line spacing gratings (Thomas 2003) and corresponds to a holographic grating produced with an aberrated wavefront via deformable mirror technology. This results in a radial change in groove density and a traditional surface of concentric hyperboloids from holography, like those used in ISIS (Beasley et al. 2004) and the Cosmic Origins Spectrograph (Green et al. 2012).

### 3.2.3. *CHESS Detector*

The cross-strip MCP detector was built and optimized to meet the CHESS spectral resolution specifications at Sensor Sciences (Siegmund et al. 2009). The detector has a circular format and a diameter of 40 mm. The microchannel plates are lead silicate glass, containing an array of 10-micron diameter channels. They are coated with an opaque cesium iodide (CsI) photocathode, which provides QE (35-15%) across the CHESS FUV bandpass (100 − 160nm). The detector operates by essentially the same principal as described for the SLICE detector, but with fewer MCPs. There are two MCPs arranged in a "chevron" configuration. During flight, the detector achieved spatial resolution of 25 µm over an 8k pixels x 8k pixels format. While spatial resolutions over these size detectors have been achieved before with cross delay line anodes, the new cross strip format is capable of much higher global count rates, providing a first flight demonstration of a higher dynamic range MCP that will be essential for future large missions that will have high photon arrival rates.

Following the launch of the CHESS-1, and in preparation for CHESS-2, we have had the 40mm cross-strip MCP refurbished at Sensor Sciences. This included the deposition of a new CsI photocathode, QE characterization, new digital output cables, and control systems functionality tests carried out in coordination with our team. A photo of the CHESS-2 detector, following refurbishment, is shown in Figure 6 (left).

### 3.3. *Suborbital Imaging Spectrograph for Transition region Irradiance from Nearby Exoplanet host stars (SISTINE)*

At present, we are in the design phase of the SISTINE payload. SISTINE is an *f*/33 imaging spectrograph comprised of an *f*/16 classical Cassegrain telescope and a magnifying spectrograph. The system is designed for R ≈ 10,000 spectroscopy across the 100 − 160 nm bandpass with imaging performance between 0.5 − 2.0″ depending on the field position and wavelength (optimized for R ≈ 11,900 and Δθ ≈ 0.5″ at Lyα; Figure 7). SISTINE employs new advanced Al+LiF ("eLiF") coatings developed at GSFC (Quijada et al. 2014) and is projected to deliver a peak effective area of $A_{eff} ≈ 170$ cm$^2$ at Lyα (Section 4.3), enabling high-sensitivity, moderate-resolution, astronomical imaging spectroscopy across this bandpass for the first time.

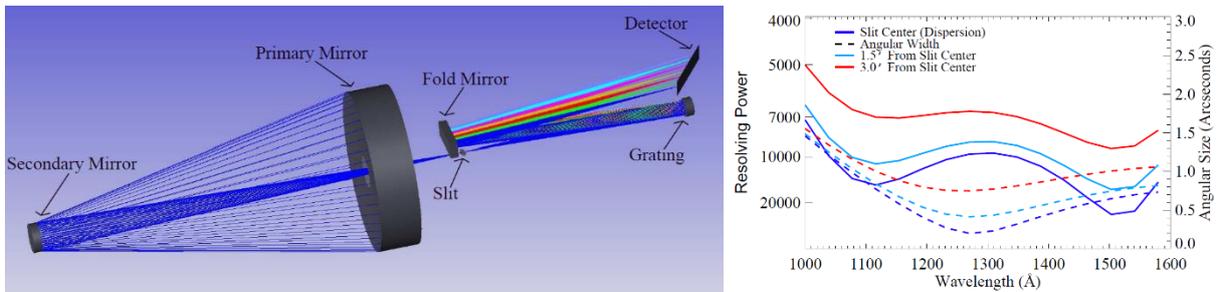

Figure 7. At left, we show a raytrace of the SISTINE payload. The telescope is a 0.5-m primary Cassegrain design. Ar right, we show performance projections for SISTINE. The solid lines show the predicted spectral resolving power (left axis) and the dotted lines show the imaging performance as a function of wavelengths and field position (in arcsec, right axis).

| Instrument Parameter | SISTINE | STIS G140M |
|---|---|---|
| Spectral Resolving Power | 10,000 | 10,000 |
| Total Spectral Bandpass | 100 – 160 nm | 114 – 174 nm |
| Spectral Bandpass per Exposure | 60 nm | 5 nm |
| Number of Exposures to Cover Spectral Bandpass | 1 | 12 |
| Imaging Field-of-View | 360″ | 28″ |
| Spectrograph Throughput | 14.1% | 1.2% |

Table 3. Performance comparison between SISTINE and the G140M long-slit imaging mode on HST-STIS. SISTINE offers $\geq 10 \times$ performance enhancement in bandpass, field-of-view, and throughput for $R = 10^4$ imaging spectroscopy in the far-UV.

| Telescope | | Spectrograph | System |
|---|---|---|---|
| Cassegrain Focal Ratio : $f/16$ Slit Dimensions: 5.15" x 6' | | Bandpass (Å): 1000 to 1600 R ≈ 9000 - 11900 | Focal Ratio : $f/33$ Effective Area : 171.7 cm$^2$ |
| Primary Mirror | Grating | Secondary Mirror | Fold Mirror (W x H): 140 x 40 mm |
| Radius (mm): 2500 OD (mm): 500 Figure: Parabola Coating: eLiF | Radius (mm): 998 Figure: Hologram Blaze λ: 1200 Å Coating: eLif | Radius (mm): -498 OD (mm): 86 Eccentricity: 1.37 Coating: eLiF | Detector |
| | | | Type: Windowless Borosilicate CsI Cross-delay line MCPs Size: Two 110 x 40 mm |

Table 4. Preliminary instrument specifications for the SISTINE sounding rocket payload.

### 3.3.1. *SISTINE Telescope*

SISTINE's $f/16$ telescope consists of a 0.5 m diameter (500 mm) parabolic primary mirror and a 86 mm convex hyperbolic secondary (Figure 7, Table 4). The detector-area limited field-of-view is 6′. The primary mirror is $f/2.5$, and will be fabricated into a single-arch fused-silica substrate by Nu-Tek Corporation with a 1mm edge chamfer to facilitate mounting. The primary has decenter and tilt tolerances of ± 1 mm and 0.2º, respectively; the optomechanical design will be capable of compensating while maintaining system performance. The 0.5 m primary can be accommodated within a standard WFF 22-inch diameter rocket skin.

The secondary mirror will have a −498 mm radius of curvature and will also be fabricated from fused silica. We anticipate that the most challenging aspect of the optical assembly will be the relative x-y centering of the primary and secondary mirrors (the tilt and z (focus) can be readily controlled by machine tolerances and standard 3-point mounts). We will incorporate the expertise of the optical engineering facilities at the University of Colorado (LASP) for the design and metrology of the optical instrument. Following telescope alignment, we will verify the environmental stability of the SISTINE telescope with a vibration test at Cascade TEK in Longmont, CO. The imaging performance will be quantified before and after vibration in our dedicated calibration facilities at the University of Colorado (described in Section 4).

### 3.3.2. *SISTINE Spectrograph*

The SISTINE spectrograph is fed through a long-slit aperture (projected angular dimensions 5″ × 6′) at the telescope focus. The slit mount will be angled at 45º about the slit axis, polished, and optimized for visible reflectivity; the reflected image of the telescope field will be folded into a visible-light aspect camera to enable real-time maneuvering feedback during the flight. The light passing through the slit is dispersed by an aberration-corrected holographically-ruled spherical grating 745 mm from the slit toward the forward end of the rocket.   The spherical figure will have a radius of curvature of 998 mm and is blazed to a peak efficiency at 120 nm, with an effective ruling density of 1278 grooves mm$^{-1}$.   The grating substrate is fused silica and will be fabricated by Precision Asphere, while the holographic ruling will be carried out by J-Y Horiba.

The use of eLiF coatings permits the inclusion of a fold mirror without a crippling loss of effective area. This allows the grating-detector distance to be larger than the slit-grating distance while fitting within the rocket envelope. This creates additional magnification in the spectrograph, enabling sufficient imaging resolution to reliably separate Lyα emission from the astronomical target from that introduced by airglow (and to angularly resolve nearby binary star systems).   The long focal length also allows us to control aberrations and provide moderate resolution spectroscopy and imaging across large spectral (60 nm) and spatial regions (6′) in a single observation (see Table 3 for a comparison of the instrument capabilities of SISTINE and HST-STIS' G140M medium resolution imaging spectroscopy mode).   The fold mirror will be on a 3-axis controllable mount so that the z-position and tilt of the fold mirror can compensate for fabrication and alignment errors in the grating figure, ensuring that the system can be assembled to spec.

The dispersed and imaged beam is recorded at the focal plane by a large format microchannel plane detector employing a CsI photocathode and a cross delay line (XDL) anode readout.  The XDL MCP will consist of two 110 mm × 40 mm segments with a common anode, and will be purchased from Sensor Sciences, building on their extensive heritage of UV detectors for NASA's space missions. The size of these detectors is too large to utilize the same cross strip technology employed by CHESS, therefore the global count rate of SISTINE will be limited to a few x $10^5$ Hz. The detector will be mounted at a 30° angle relative to the incident beam in order to achieve the best focus and resolution while maintaining a flat focal plane, which minimizes costs. The XDL anode provides a photon-counting event list to enable the time-resolved spectroscopy described in Section 2.2.  The large format, higher open area afforded by the new ALD borosilicate MCPs (10 – 15% larger than conventional MCP glass; Siegmund et al. 2011), and the optimized pore pitch (up to another 10% efficiency gain) support the COR technology priority for the development of higher-QE, large format, photon-counting UV detectors.   SISTINE will provide the first flight test for these detector technologies in an end-to-end astronomical instrument.

## 4. Payload Calibration, Laboratory and Flight Performance

Absolute and relative calibrations are essential for the full characterization of a new flight technology and hence to quantitative astronomical measurements.  We carry out testing, calibration, and characterization of the detector system in the vacuum test facilities at the Center for Astrophysics and Space Astronomy (CASA), on the campus of the University of Colorado at Boulder.   This facility utilizes several component- and space-flight payload sized vacuum chambers for instrument testing in simulated flight conditions (these facilities have a long history of supporting ultraviolet astrophysics missions including testing of the FUSE spectrograph and HST-COS components). We use this facility to measure detector quantum efficiencies, intrinsic detector resolution, and the flux dynamic range (count rate limits) at vacuum ultraviolet wavelengths, from 400 – 3000 Å. The facility uses a combination of lamps, evacuated monochromators, and a vacuum ultraviolet collimator to provide illumination for testing.  Secondary absolute calibration standards referenced to NIST photodiodes enable absolute calibration of all subsystems and the determination of the end-to-end throughput of the instrument.  We extend this calibration to an absolute scale with photomultiplier tubes and MCPs referenced to NIST photodiodes.  UV continuum spectra allow for the construction of detector flat-field maps that can be used to correct for detector non-uniformities during the characterization process (France et al. 2002).

Mirror reflectivity and grating groove efficiency measurements are obtained from a relative comparison of an incident, unobstructed beam of incident photons with the same beam (assumed to be generated by a stable source) after reflecting/diffracting off the target optic. The wavelength dependence of these quantities is explored through the use of windowless light source with many transitions across the far-UV bandpass and an evacuated monochromator for wavelength selection when necessary.  Because we are focused on the entire far-UV bandpass covered variously by SLICE, CHESS, and SISTINE, we typically use a 65% Ar / 35% $H_2$ gas mixture into a gas discharge lamp operating at ~100 mTorr and ~300 V, thereby creating a bright, discrete emission line source in the far-UV band. In order to thoroughly characterize the flight optics, we measured the reflectivity of each component using emission lines spanning $92 - 160.8$ nm (ArII 92.0, 93.2; Ly$\beta$ 102.6; Ar I 104.8, 106.6; $H_2$ 111.5, 114.3, 116.4, 117.3, 125.4, 128.0, 140.0, 143.6, 157.5, 160.8; and Ly$\alpha$ 121.6). We have described these measurements in considerable detail for the SLICE instrument previously (France et al. 2013c), so we simply summarize them here and focus more on new measurements of the CHESS grating performance.

## 4.1. *SLICE Resolution and Effective Area*

Throughout the far-UV bandpass, the most robust wavelength calibration source are the electron-impact excited emission lines of $H_2$ (Lyman and Werner bands), HI Lyman series, and Ar I 104.8, 106.6 nm. Additional lines of N I, O I, and CO are available with the appropriate combination of target gases fed into the discharge lamp. The observed calibration spectra are compared to a synthetic electron-impact spectrum (France et al. 2010a, 2011) assuming a 400 K thermal population of $H_2$ bombarded by electrons with a mean energy of 100 eV. Using deep Ar/$H_2$ spectra ($T_{exp} \sim 10 - 20$ minutes at global count rates $300 - 600$ Hz), we were able to construct an accurate wavelength solution and demonstrate instrumental spectral resolution down to the measured spatial resolution of the Quantar MCP detector ($\Delta \approx 100$ $\mu$m). The average FWHM of the calibration emission lines was $96.1 \pm 13.9$ $\mu$m, which translates into a system resolving power of $5304 \pm 733$ or a velocity resolution of $\delta v = 56 \pm 8$ km/s. These results were confirmed through in-flight observations of the intrinsically narrow Ar I interstellar absorption lines (Figure 8, left).

These pre-flight measurements were confirmed by the in-flight spectra obtained by SLICE. The rich suite of atomic and molecular absorption features in the ISM confirms both the wavelength and resolution calculations, while a comparison of our observations of the bright B1 V star $\alpha$ Vir to stellar atmosphere models allow us to characterize the end-to-end effective area of the SLICE instrument. We modeled the IUE spectra of $\alpha$ Vir (Spica) with the TLUSTY B-star models (Lanz & Hubeny 2007), finding good agreement with the observed $116.0 - 125.0$ nm data for a model with $T_{eff} = 24,000$ K, $\log(g) = 3.33$, $Z = 0.5Z_{\odot}$, and $v_{turb} = 2$ km s$^{-1}$, scaled to an 115.0 nm flux, $F(115)$ = $8.3 \times 10^{-8}$ erg cm$^{-2}$ s$^{-1}$ Å$^{-1}$ (see Figure 8, right). A spline function was fitted to the ratio of stellar model-to-SLICE observations for 15 points relatively unobscured by strong stellar or interstellar absorption lines, creating a smooth flux calibration curve from $102.0 - 107.0$ nm, in units of [erg cm$^{-2}$ s$^{-1}$ Å$^{-1}$ / counts s$^{-1}$]. These results, based on data taken during the 36.271 UG mission, are shown in Figure 8. The peak effective area of SLICE was approximately 1.3 cm$^2$ at 104.2 nm (Figure 8, center).

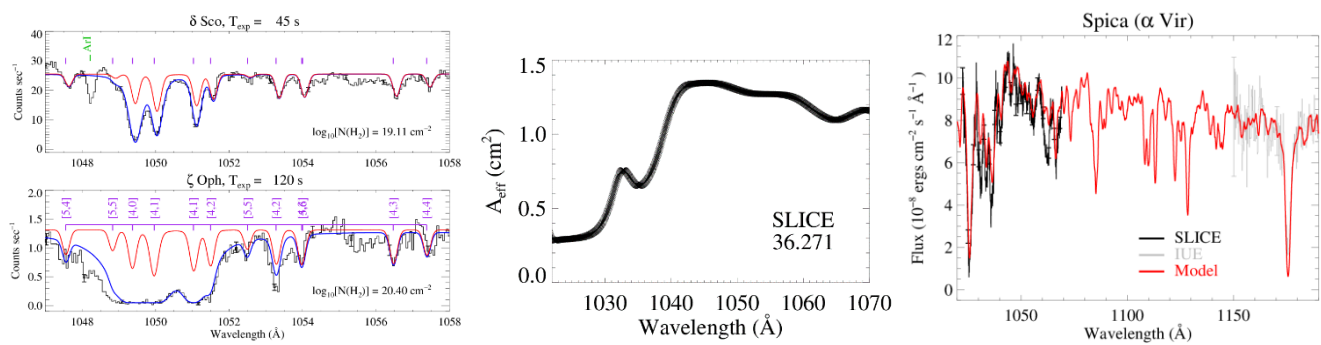

Figure 8. (left) Raw data from observations of δ Sco and ζ Oph obtained during the 36.271 UG / SLICE mission. Interstellar absorption lines are immediately detectable without significant post-processing. (center) The effective area curve of the SLICE instrument, as derived from a comparison between flight data and stellar atmosphere models. (right) Flux calibration was performed by modeling the spectrum of α Vir using the TLUSTY B-star database and normalizing the stellar model to IUE spectra. The stellar continuum flux ratio between SLICE and the model was used to create a smoothly varying flux calibration transfer curve that was applied to all four targets (figure adapted from France et al. 2013c).

## 4.2. *CHESS Resolution and Effective Area*

Alignment procedures for the CHESS echelle and cross disperser have been described previously by Hoadley et al. (2014). Following spectrograph build-up, long exposures of pure hydrogen gas through the arc lamp were taken for a complete sampling of H and $H_2$ emission lines in the CHESS bandpass. A total of 10 million counts were collected in the final calibration spectrum over 4 separate exposures, the two dimensional calibration spectrum is shown in Figure 9 (left). The $100 - 160$nm spectrum is distributed top-to-bottom (in increasing wavelength) across the detector face.

To understand the distribution of $H_2$ lines relative to Lyα, a series of order extractions were made for orders which showed several emission features with high S/N over the entire bandpass to preliminarily identify the wavelength coverage of each order. Spectral identifications were made by comparing the calibration data with $H_2$ arc lamp fluorescence models and vacuum $H_2$ emission identifications (Roncin & Launay 1994). Each order was physically located in pixel-space, and order width contribution to the extracted 1D spectrum was under-sampled to avoid light contamination from adjacent orders. In most cases, the order widths were taken to be 10 pixels wide in $2048 \times 2048$ pixel space. Gaussian line fits were then carried out on every emission line in the final one-dimensional co-added spectrum, using an interactive fitting routine developed originally for HST-COS and subsequently adapted for analysis of SLICE and CHESS laboratory data.

Overall, the narrowest $H_2$ features are found in the $150 - 160$nm range, and have a FWHM corresponding to R ~ 70,000 $\lambda/\Delta\lambda$. For shorter-wavelength $H_2$ emission features, R ~ $20,000 - 40,000$ $\lambda/\Delta\lambda$ is typical for lines shortward of $\lambda = 130$nm. In this way, both the wavelength solution and the instrumental line-spread-function are determined. Figure 9 (right) shows the resulting 1D extracted spectrum, including the wavelength solution. There are gaps between order segments, caused by the out-of-specification performance of the echelle grating on CHESS-1. Because the echelle has a larger groove density than designed for the detector collecting plane, all orders lose a portion of their free spectral range to the edges of the detector. The red end of the bandpass loses more wavelength coverage because the free spectral range increases as wavelength increases, and the circular format of the detector inherently reduced collecting area for the longer wavelengths. CHESS-2 will experience similar gaps as moved to a higher groove density to enable groove efficiencies that meet the minimum flight requirements (as described above).

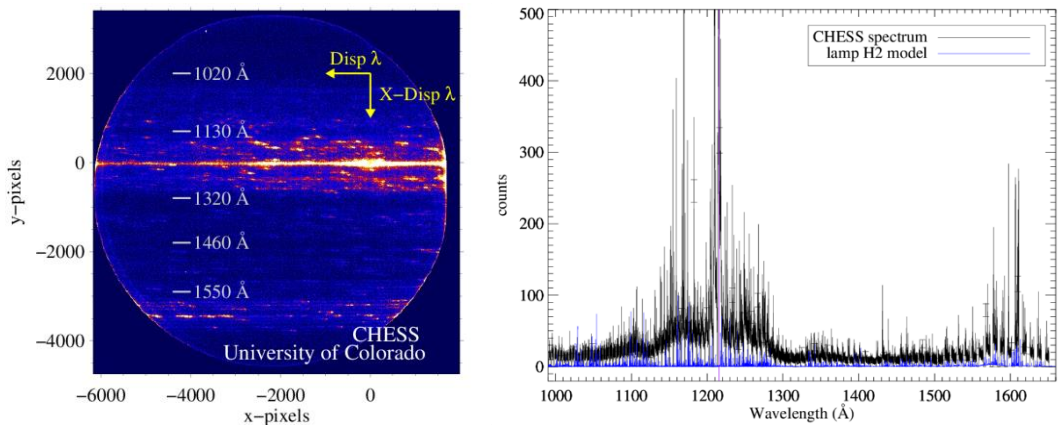

Figure 9. Calibration data from the CHESS instrument (*left*), demonstrating R ≈ 70,000; the orders are separated top-to-bottom, each order has 5 − 7 Å of free spectral range. The extracted one-dimensional spectrum is shown at right (Hoadley et al. 2014). A grating mount shift during launch on 36.285 reduced the instrumental resolution and efficiency (see Figure 14 below).

Component level efficiencies for the CHESS-1 instrumental configuration are given in Hoadley et al. (2014). As noted in Section 3.2, the 40mm diameter cross-strip anode MCP was refurbished following CHESS-1, including the deposition of a fresh photocathode. This has increased the quantum efficiency (DQE) of the detector to 27% at 99nm, 39% at 113nm, and 26% at 120nm. The final groove design for the electron-beam etched echelle grating for CHESS-2 has been identified and sample grating tests show groove efficiencies peaking at ~6% across the $92 - 160$nm far-UV bandpass. A comparison of groove efficiency curves are shown in Figure 10.

We can express the instrumental effective area as: $A_{eff}[CHESS] = A_{geom} \times G_{ech}(\lambda) \times G_{Xdisp}(\lambda) \times DQE(\lambda)$, where $A_{geom}$ is the geometric collecting area, including the grating tilt to accommodate the $67°$ angle-of-incidence and the losses owing to the physical size of the mechanical collimator structure. $A_{geom} = 27$ cm². $G_{ech}(\lambda)$ and $G_{Xdisp}(\lambda)$ are the total grating reflectivities for the echelle and cross-disperser, respectively, and are the product of the groove efficiency and the reflectivity of the coated substrate material ($G_{ech}(\lambda) = E_{ech}(\lambda) \times R_{ech}(\lambda)$); each measured in the calibration facilities at Colorado in 2015. $E_{ech}(\lambda) = 10\%$ and $R_{ech}(\lambda) = 50\%$ for the JPL e-beam etched grating. An analogous expression can be created for the cross-disperser, with $E_{Xdisp}(\lambda) = 40\%$ and $R_{Xdisp}(\lambda) = 50\%$. Taken together, the CHESS instrument achieves peak $A_{eff} = 0.1$ cm², which stays flat to within ~30% over the CHESS bandpass. The target for CHESS-2, the bright B0.5 V star ε Per, has a 120 nm continuum flux of $1 \times 10^{-8}$ erg cm⁻² s⁻¹ Å⁻¹, we estimate S/N = 10 per resolution element (at $R = 100{,}000$) during the nominal 300 second exposure time for the Black Brant IX vehicle we will launch on during the CHESS-2 (36.297 UG) mission.

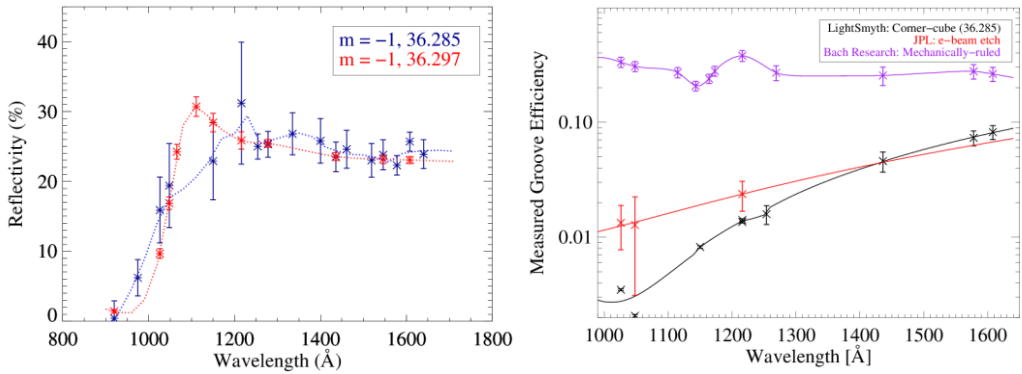

Figure 10. At left, CHESS cross-disperser reflectivity performance (note this is the total component reflectivity $G_{ech}(\lambda) = E_{ech}(\lambda) \times R_{ech}(\lambda)$, see text). The Al+LiF coating shows no signs of degradation following 36.285 UG integration, launch, travel, and approximately nine months of dry storage. At right, the echelle groove efficiency performance for the Lightsmyth dry-etched grating, the electron-beam etched grating from JPL, and a sample 52.9 groove mm⁻¹ mechanically ruled grating from Bach Research.

### 4.3. *SISTINE Effective Area Calculations*

The SISTINE payload is still in the design phase, but we can provide an estimate of the expected sensitivity of the instrument. The SISTINE effective area can be written: $A_{eff}[SISTINE] = A_{geom} \times R(\lambda)^3 \times G(\lambda) \times DQE(\lambda)$, $A_{geom}$ is the geometric clear area of the telescope (1905 cm², with > 95% slit transmission), $R(\lambda)$ is the near-normal incidence reflectivity of enhanced Al+LiF coatings from GSFC (conservatively taken as 80%), $G(\lambda)$ is the product of the groove efficiency and the reflectivity of the coated substrate material ($G(\lambda) = E(\lambda) \times R(\lambda)$), where we expect $E(\lambda)$ ~ 55% (based on our group's experience with similar gratings for HST-COS) and $R(\lambda)$ ~ 80%. The SISTINE MCP detector is expected to have a peak DQE ~ 40%. Taken together, the SISTINE instrument achieves peak $A_{eff} = 171$ cm² at 120 nm, with a minimum of 140 cm² at 160 nm.

For the first flight of SISTINE, we plan to observe the active M dwarf AD Leo ($d = 4.7$ pc). AD Leo is the most common proxy star used in calculations of terrestrial planet atmospheres around M dwarfs, although most authors must currently rely on a disjointed combination of archival *IUE* data filled in at specific wavelengths by *HST*. Assuming the projected effective area, SISTINE achieves S/N = 6.6 at O VI and 7.7 at Lyα in a 300 second sounding rocket exposure. It is important to note that the Lyα S/N estimate includes the subtraction from the geocoronal contribution; imaging spectroscopy is essential for the measurement of Lyα, without a spatially resolved airglow spectrum "above and below" the stellar spectrum, the star's Lyα spectrum cannot be effectively separated from the geocoronal background (France et al. 2012). This is one of the imaging spectrograph science drivers for SISTINE.

## 5.    SLICE and CHESS Flight Data Science Summary

### 5.1.    *NASA/CU 36.271 UG – SLICE*

SLICE was launched aboard a Terrier-Black Brant IX sounding rocket from Launch Complex 36 at the White Sands Missile Range at 02:00 MDT on 21 April 2013 as part of NASA mission 36.271 UG. The payload aperture, a vacuum-sealed door, opened at T+72 seconds, at which point the payload passively evacuated for 30 seconds prior to detector high-voltage turn on at T+102 seconds. The first target was acquired by science team operator command uplink to the on-board attitude control system. Target 1 (η Uma) was placed in the spectrograph aperture by referencing real-time video of the slit jaw acquired by a Xybion intensified camera and relayed to the command center via telemetry. The ST5000 star tracker/experiment alignment was then re-centered at the nominal ``on-target'' pointing, and subsequent maneuvers between targets required only minor peak-up adjustments. SLICE reached apogee at T+286s at an altitude of 309 km. The payload shutter door was closed at T+507 seconds and reentry began near T+550s. The parachute deployed at T+650s and touchdown occurred at approximately T+915 seconds, about 89 km to the northwest of the LC 36 launch site.

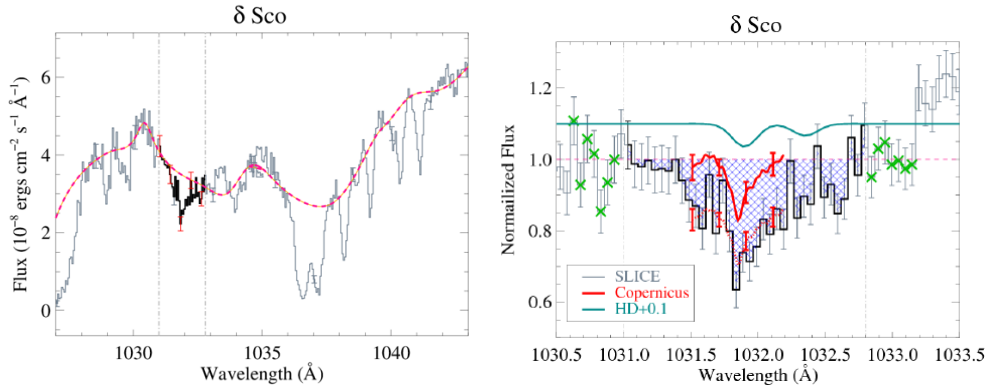

Figure 11. At left, flux calibrated spectrum of δ Sco from the 36.271 UG / SLICE mission (gray). The dashed pink line shows the spline-fit continuum normalization and the thick black region is the extent of the O VI line profile. At right, a blow up of the normalized O VI absorption line region, with the normalized line profile from Copernicus shown in the solid red line. The dotted red line shows that the Copernicus spectral shape matches that observed by SLICE, however the narrow bandpass of the Copernicus scan prevents a reliable normalization. Therefore, the SLICE data indicate that there is substantially more equivalent width in the O VI line (i.e., higher column density or multiple components) than inferred from the Copernicus data alone.

The SLICE detector acquired data for ≈ 400s during the flight, with acquisitions divided into "on-target" times when the stars were in the spectrograph aperture and the pointing was stable. The detector was on during the slews between targets and these period were extracted from the telemetry data to establish the in-flight background rate. This resulted in exposure times of $T_{exp}$ = 45s, 30s, 45s, and 120s for η Uma, α Vir, δ Sco, and ζ Oph, respectively. We achieved signal-to-noise ratios (S/Ns) of > 30 per pixel on the first three targets and S/N ~ 12 per pixel over the middle of the SLICE bandpass (104.5 – 106.0 nm) on ζ Oph. The wavelength and flux calibration curves described above were applied to the flight data to produce final, calibrated one-dimensional spectra for analysis.

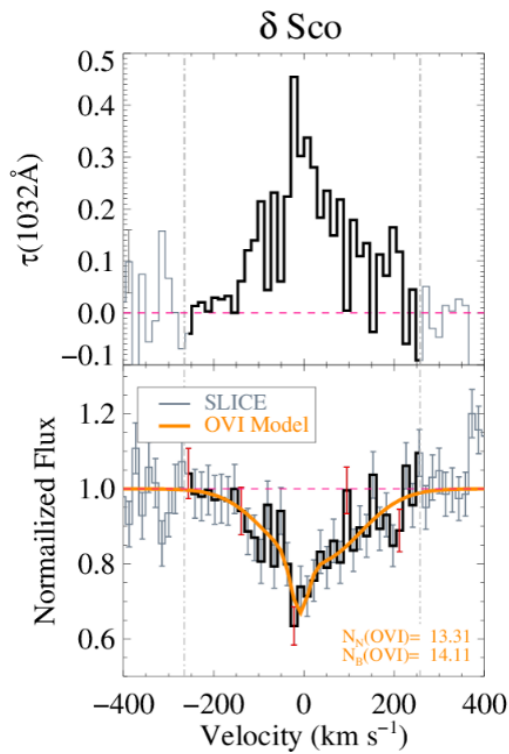

Figure 12. Model fits to the column density of the O VI 1032 line profile based on the optical depth method (Jenkins 1978). The SLICE data are consistent with a two component model, dominated by a hot, higher column density component that was not apparent from Copernicus data alone.

Figure 11 (left) shows the flux calibrated spectrum of δ Sco, with a spline fit to the stellar continuum that is used as the normalization function for absorption line analysis. The spline function is normalized to the average continuum level on either side of the O VI absorption line (indicated by the green Xs in Figure 11b). For the example presented here, we highlight the O VI 103.2 nm absorption line thought to originate in a hot interface near δ Sco, either associated with a wind-blown bubble caused by δ Sco or possibly an interface between the local bubble and the local interstellar medium in the Upper Scorpius star-forming cloud. An archival Copernicus spectrum of δ Sco is shown in red for comparison (York 1974; Jenkins & Meloy 1974). Results from SLICE show that when the improved continuum normalization resulting from a larger spectral bandpass is included, the O VI absorption towards δ Sco has considerably higher equivalent width than was deduced from the continuum placement determined from Copernicus data alone (Figure 11, right). Shifting the continuum level up or down within the RMS scatter in the green Xs leads to a systematic uncertainty of ~20% in the determination of the O VI column density (a formal analysis of the error distribution is ongoing). Removing the potential contamination from HD (< 13% of the total O VI equivalent width, based on the measured $H_2$ column density, assuming a $HD/H_2$ ratio = $10^{-5}$), we can use the optical depth methodology developed by Jenkins (1978; Figure 12) to estimate the column density and temperature limits for the broad and narrow O VI components seen in the SLICE data, [log(N(OVI)) , log(T(OVI))] = [13.31, < 5.69] and [14.11, < 7.25] for the narrow and broad components, respectively. The narrow component is consistent with the original Copernicus column density, log(N(OVI)) = 13.26 (Jenkins 1978).

SLICE also made a thorough characterization of the molecular content of the sightlines towards all four target stars, showing a break in the $H_2$ column density at the edge of the local bubble (d ~ 100 pc). The description of the data analysis for the molecular properties are given in France et al. (2013b) and we show the extracted molecular excitation diagrams in Figure 13. A mid-IR image of the region around ζ Oph is shown at right in Figure 13, demonstrating the interaction this high-velocity star with the ambient interstellar material through which it is moving. We have compared the $H_2$ properties with archival Copernicus data to examine potential time-evolution in the interstellar column densities. We conclude that the bulk molecular material (in the lowest rotational states, J″ = 0 and 1) are consistent with previous results. However, the SLICE observations find N(J″) values 0.3 − 1.0 dex higher for J″ = 2 − 6, with very similar excitation temperatures ($T_{exc}$ = 324 for Copernicus vs. 350 for SLICE K) and b-values ($b$ = 3.8 for Copernicus vs. 4 km s⁻¹ for SLICE). A direct comparison of the SLICE equivalent width with those presented in Spitzer et al. (1974) show equivalent widths ∼ 15 − 25% larger in the SLICE data, approximately the expected increase in equivalent width (4 − 37%) that would be expected from the larger SLICE column densities derived by profile fitting (France et al. 2013b).

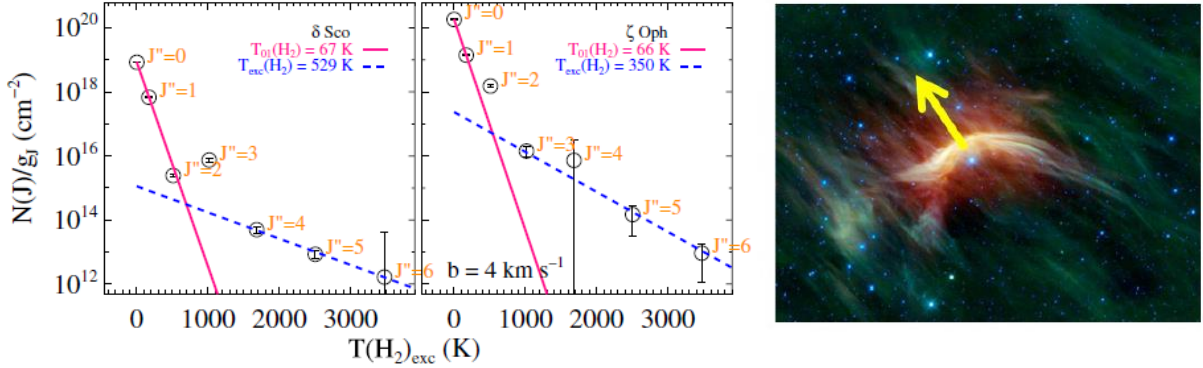

Figure 13. For the two stars beyond the ∼ 100 pc boundary of the local bubble, we used an automated $H_2$ absorption line fitting routine to measure the molecular columns as a function of rotational excitation level and characterize the kinetic temperature of the gas (*left and center*). We detect small variations in the high-J column densities on the ζ Oph sightline relative to the ∼40 year old Copernicus measurements that we tentatively attribute to the high space velocity of the star as it moves into an ambient interstellar cloud *(right)*.

## 5.2. *NASA/CU 36.285 UG − CHESS-1*

CHESS-1 was launched aboard NASA mission 36.285 UG from WSMR on 24 May 2014 at 01:35am MDT atop a two-stage Terrier/Black Brant IX vehicle. Overall, the mission was a comprehensive success and the instrument successfully collected data, approximately 300s of on-target time on α Vir. The count rate from α Vir was lower than expected, but stellar absorption features in the echellogram are readily observed. Additionally, count rate contribution from air glow contamination was lower than expected, which suggests that at least one of the optics and/or optical coatings were affected sometime between pre-flight calibrations and launch. Post-flight calibrations have suggested that the echelle mount shifted during powered ascent, reducing the throughput and resolution of the instrument during flight. Despite the reduced data quality from this first flight, several interstellar and photospheric features are readily observed in the data. Stellar features observed in the echellogram during flight include the stellar Lyα, Si III (120.6nm), and C III (117.5nm) absorption lines. Additional post-processing ongoing to extract further features, but evidence of Si II, Mg II, and many blue-ward absorption features (< 115nm) are present. Figure 14 (left) shows a wavelength calibration spectrum in the 115 − 120 nm region and an extracted, preliminary profile fit of one absorption feature from the same portion of the echellogram (Figure 14, right). As discussed, the echelle grating is being fabricated using a completely new technique for CHESS-2 and the echelle grating mount is being redesigned to include additional rigidity at the time of writing.

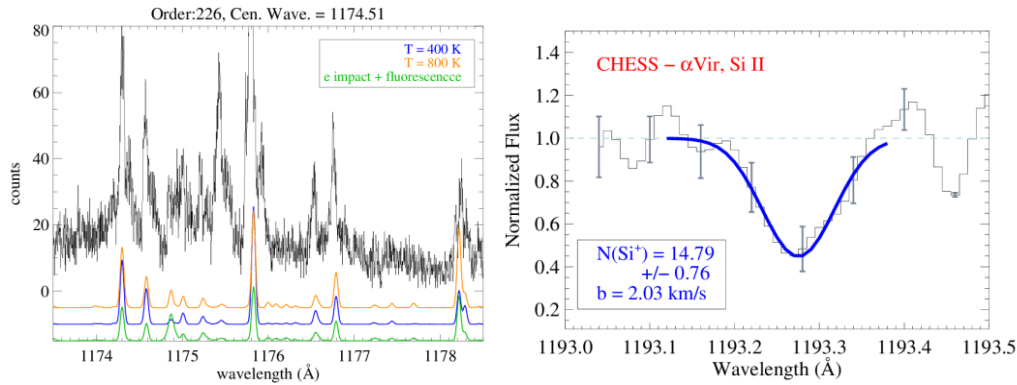

Figure 14. At left, we show a laboratory electron-impact $H_2$ spectrum acquired with CHESS during pre-flight calibrations (black histogram). The colored spectra are model $H_2$ emission spectra created for the densities and temperatures expected in the discharge lamp environment. Errors in the grating fabrication (as described in the text) limit the instrumental resolving power to $R \sim 40,000$ in the $1100 - 1200$ Å region for the CHESS-1 instrumental configuration. At left, we show one of the low-ionization ISM lines (Si II 1193) observed during the flight of CHESS-1 in May 2014. The blue line is a model fit to the line profile to measure the Doppler b-value and the log of the column density of $Si^+$ (figure adapted from Hoadley et al. 2014), consistent with archival HST-GHRS spectra of the low-ionization atomic sightline towards α Vir.

## Acknowledgments

We acknowledge the hard work and dedication of the NASA Wallops Flight Facility/NSROC payload team, the Physical Sciences Laboratory at New Mexico State University, and the Navy team at WSMR that support all Colorado astronomy payloads. We appreciate Ted Schultz for assistance with the SLICE electronics package. This work was supported by NASA grants NNX10AC66G and NNX13AF55G to the University of Colorado at Boulder. The authors thank Drs. Charles Danforth, Ed Jenkins, Seth Redfield, Aki Roberge, John Stocke, and Jason Tumlinson for their scientific and programmatic support for the CU UV rocket program.